\documentclass[10pt,a4paper,5p,sort&compress]{elsarticle} % review
\usepackage{amsfonts}
\usepackage{amsmath}
\usepackage{xcolor}
\usepackage{hyperref}
\hypersetup{
    colorlinks,
    linkcolor={red!50!black},
    citecolor={blue!50!black},
    urlcolor={blue!80!black}
}
\usepackage{graphicx}
\usepackage{url}

\journal{Nuclear Instruments and Methods in Physics Research Section A}

\usepackage[singlelinecheck=false]{caption}
\begin{document}
\begin{frontmatter}
\title {Dual Position Sensitive MWPC for tracking reaction products at {\sc VAMOS}++}

%\author{Authors ...} 
\author{M.~Vandebrouck} 
\author{A.~Lemasson\corref{cor1}} 
\ead{lemasson@ganil.fr}
\cortext[cor1]{Corresponding author}
\author{M.~Rejmund\corref{cor2}}
\author{G.~Fremont}
\author{J.~Pancin}
\author{A.~Navin}
\author{C.~Michelagnoli}
\author{J.~Goupil}
\author{C.~Spitaels}
\author{B.~Jacquot}
\renewcommand{\topfraction}{0.85}
\renewcommand{\bottomfraction}{0.85}
\renewcommand{\textfraction}{0.05}
\renewcommand{\floatpagefraction}{0.99}

\address{GANIL, CEA/DSM - CNRS/IN2P3, Bd Henri Becquerel, BP 55027,
  F-14076 Caen Cedex 5, France}
\begin{abstract}
The  characteristics  and performance  of  a  Dual Position  Sensitive
Multi-Wire  Proportional  Counter   (DPS-MWPC)  used  to  measure  the
scattering  angle, the  interaction  position on  the  target and  the
velocity of  reaction products detected in the  {\sc VAMOS}++ magnetic
spectrometer,  are  reported.  The  detector  consists  of  a pair  of
position sensitive  low pressure MWPCs  and provides both  fast timing
signals, along with  the two-dimensional position coordinates required
to define  the trajectory of the reaction  products.  A time-of-flight
resolution  of   $305(11)$~ps  (FWHM)  was   measured.   The  measured
resolutions  (FWHM)  were  $2.5(3)$~mrad  and $560(70)~\mu$m  for  the
scattering angle and the interaction point at the target respectively.
The subsequent improvement of the  Doppler correction of the energy of
the $\gamma$-rays,  detected in  the $\gamma$-ray tracking  array {\sc
  AGATA}  in coincidence  with  isotopically identified  ions in  {\sc
  VAMOS}++, is also discussed.


\end{abstract}
\end{frontmatter}

\section{Introduction}
Nuclear  reactions  at  energies  around  the Coulomb  barrier  are  a
powerful   tool  to  investigate   nuclear  structure   and  dynamics.
Heavy-ion  fusion  reactions in  conjunction  with large  $\gamma$-ray
detector arrays  have been used  to understand behaviour of  nuclei at
the  extremes  of  angular  momentum~\cite{Wang2013Yb}.   Isotopically
identified fission fragments, from fission induced by heavy-ion fusion
or  transfer, have  been  recently used  to characterize  neutron-rich
nuclei at large  angular momentum~\cite{Navin2014,Nav14Yb} and also to
study     the    multidimensional     facets     of    the     fission
process~\cite{Caamano2013,Caamano2015}.   More  peripheral collisions,
around the Coulomb  barrier, have also played a  role in understanding
the mechanism  of nuclear transfer~\cite{Corradi2009}  and to populate
nuclei far away from stability ~\cite{Bhattacharyya2008,Watanabe2015}.
Direct  reactions, in particular  using radioactive  ion beams,  are a
powerful           tool           to           probe           nuclear
shell-evolution~\cite{Catford2010,Fernandez2011}.

At   beam  energies   around   the  Coulomb   barrier,  the   isotopic
identification (mass  ($A$) and atomic  ($Z$) numbers) of  the various
reaction  products is challenging  for medium  mass and  heavy nuclei.
This is especially  required when the nuclei of  interest are produced
with small production cross-sections among a large number of different
reaction  products.    Isotopic  identification  can   be  efficiently
achieved using a large  acceptance spectrometer combined with suitable
detection  system.   In  the  last decade,  several  large  acceptance
spectrometers       like       {\sc       VAMOS}++~\cite{Pul08,Rej11},
PRISMA~\cite{Beghini2005}                                           and
MAGNEX~\cite{Cunsolo200248,Cunsolo2002216}     were     built.     The
performance  of such  a  spectrometer depends  on  the ion  trajectory
reconstruction method  and can be further improved  with the knowledge
of the interaction  position on the target. The  direct measurement at
the  entrance of  the spectrometer  of the  scattering angles  and the
interaction  point  on  the  target  will  result  in  less  stringent
conditions on optical properties of the beam.

The PRISMA  and VAMOS++ spectrometers  have been coupled with  a large
$\gamma$-ray    detector    array    like    CLARA~\cite{CLARA}    and
EXOGAM~\cite{Sim00}   respectively   for   probing   spectroscopy   of
neutron-rich    nuclei~\cite{Bhattacharyya2008,    Valiente-Dobon2009,
  Ljungvall2010,  Montanari2011}.   The advent  of  new generation  of
$\gamma$-ray      tracking      arrays     AGATA~\cite{AGATA}      and
GRETINA~\cite{Paschalis2013} led  to an improved  determination of the
first interaction point  in the detector and also  allowed to increase
their  operating rate  with larger  $\gamma$-ray  multiplicities.  The
increased  granularity results  in an  improved Doppler  correction of
$\gamma$-rays emitted in flight,  provided that the velocity vector of
the  recoiling  ion  is  measured  with  sufficient  precision  on  an
event-by-event basis.  Typically,  resolutions, in scattering angle of
the ion and definition of  the interaction point at the target, better
than $1^\circ$  and 1~mm respectively  are required so that  the final
Doppler corrected  $\gamma$-ray energy  resolution is limited  only by
pulse shape analysis and $\gamma$-ray tracking capabilities.

For   measuring   the   velocity    vector   at   PRISMA,   a   single
Micro-Channel-Plate   detector   placed  at   the   entrance  of   the
spectrometer~\cite{Montagnoli2005},   provided   the   two-dimensional
position and  a timing  signal for the  reaction products,  assuming a
point like beam spot on the  target.  A large area MWPC located in the
focal plane  provided the  two-dimensional position and  timing signal
for   the  time-of-flight  measurement~\cite{Beghini2005}.    At  {\sc
  VAMOS}++, start  and stop detectors  were used, at the  entrance and
focal plane  of the spectrometer~\cite{Rej11}.   The scattering angles
at the  target were obtained  from a reconstruction method  which used
the  trajectory of  the  ions measured  at  the focal  plane by  drift
chambers~\cite{Rej11,Pul08}.   The  resulting  angular resolution  was
sufficient  for  the Doppler  correction  given  the relatively  large
angular opening  of the electrical  segmentation of the  EXOGAM clover
detectors~\cite{Sim00}.  However, it  is insufficient for the superior
angular resolution of the AGATA $\gamma$-ray tracking array.

With the  above motivations, a new Dual  Position Sensitive Multi-Wire
Proportional  Counter   (DPS-MWPC),  providing  time   information,  a
measurement  of the  scattering  angle and  interaction  point on  the
target,  was developed.  Here the  characteristics and  performance of
this new detector are reported.

\begin{figure*}[t]
\begin{center}
\includegraphics[width=\textwidth]{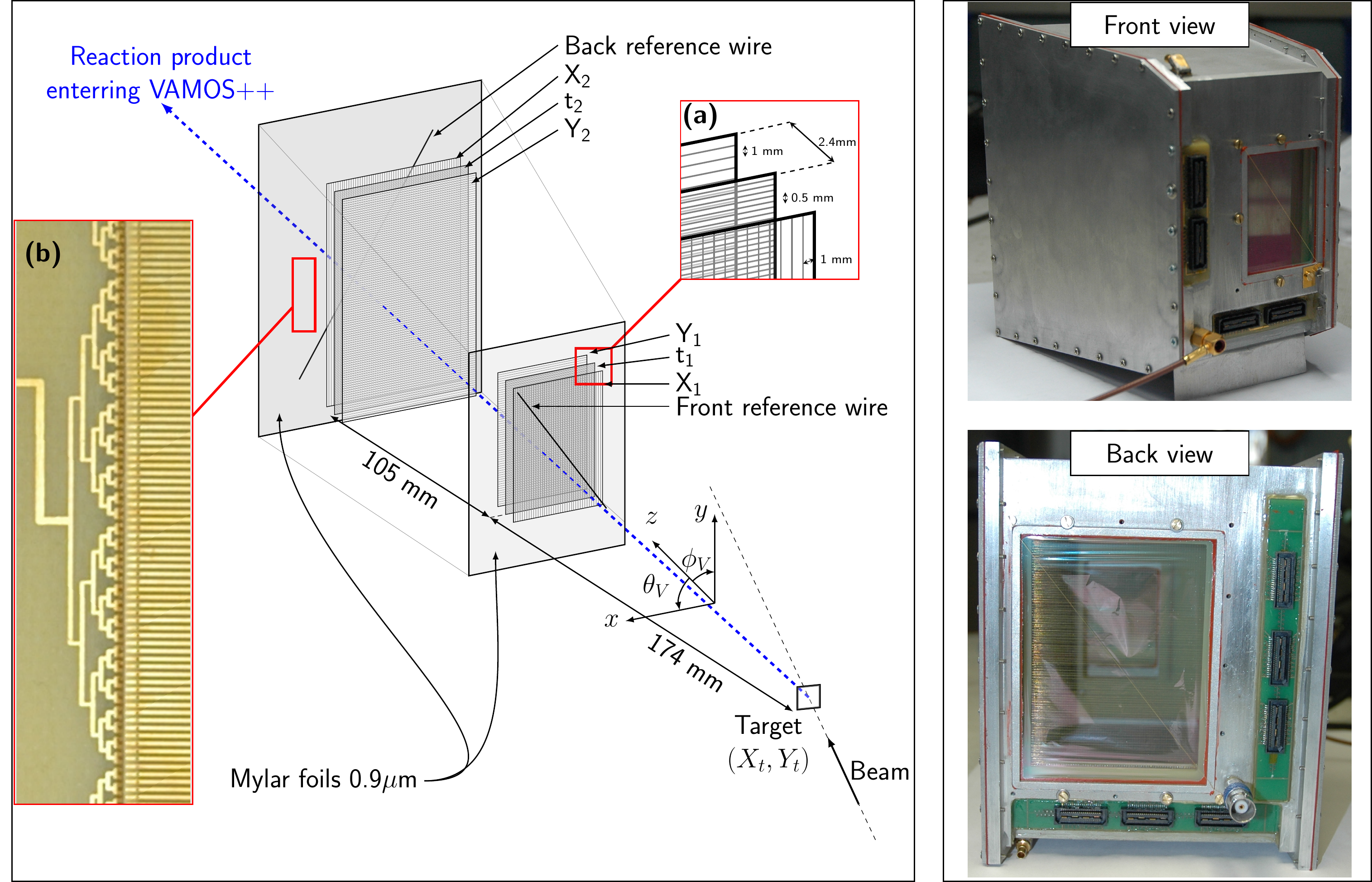}
\caption{\label{fig:fig1} (Color  online) Left: Schematic  view of the
  DPS-MWPC detector assembly.   The detector is composed of  a pair of
  position sensitive  MWPC.  Inset (a)  : Expansion of the  three wire
  planes illustrating  the wire plane orientation  and spacings. Inset
  (b): Picture of  the equivalent path length routing  of a time plane
  signal.  Right: Front and back views of the detector assembly.}
\end{center}
\end{figure*}

\section{Detector description}
The  purpose of  the present  detector is  to detect  low  energy ions
produced in  reactions at energies  around the Coulomb barrier  in the
vicinity of magnetic fringe fields  and close to target bombarded with
intense heavy ion beams.   Hence, low pressure Multi-Wire Proportional
Counters     (MWPC)~\cite{Breskin1977,Breskin1979,Breskin1982}    were
chosen.   The  new  detector  assembly  consists  of  a  pair  of  two
dimensional  position sensitive  MWPC in  a common  gas  volume placed
between  the reaction  target and  the entrance  of the  {\sc VAMOS}++
spectrometer.  A schematic  view of the detector assembly  is shown in
Fig.~\ref{fig:fig1}.   The front and  back MWPC  have active  areas of
$40\times61$~mm$^2$   and   $65\times93$~mm$^2$  respectively.    Each
detector  is composed  of  three electrodes:  a  central cathode  that
provides  a  time signal  ($t_{1,2}$)  and  two orthogonally  oriented
anodes  wire  planes  ($X_{1,2}$,$Y_{1,2}$)  (Fig.~\ref{fig:fig1}(a)).
The  cathodes  of the  two  MWPCs are  separated  by  105~mm, and  the
detector assembly is  placed 174~mm downstream of the  target.  Such a
geometry allows the  detector to cover the full  angular acceptance of
the {\sc  VAMOS}++ spectrometer ($\Delta  \theta_V = \pm  7^\circ$ and
$\Delta  \phi_V  =  \pm  11^\circ$  in  the  horizontal  and  vertical
direction  respectively).   For  each  of  the MWPC,  the  cathode  is
composed of  gold plated tungsten  wires with a diameter  of 20~$\mu$m
and  separated by  0.5~mm.  The  $X$ and  $Y$ anodes  are  composed of
20~$\mu$m gold plated tungsten wires separated by 1.0~mm.  The cathode
and   anode    planes   are   separated    by   a   gap    of   2.4~mm
(Fig.~\ref{fig:fig1}(a)).  The different spacing between the wires for
the cathode and the anodes was chosen to obtain the required avalanche
amplification gain.  To  avoid a dispersion in the  measurement of the
time signal due  to the propagation of the  signal along the different
wires, equivalent path lengths of  the cathode signal were designed so
that it  is independent of the  $Y$ position.  This  is illustrated in
Fig.~\ref{fig:fig1}(b) for  one of the  time planes used in  the MWPC.
The residual time dispersion, arising due to the varying $X$ position,
can  be corrected  on  through software  on  an event-by-event  basis.
Entrance and exit Mylar  windows with 0.9~$\mu$m thickness isolate the
gas volume.  Two gold  plated tungsten wires (100~$\mu$m diameter) are
placed diagonally  1~cm up and downstream  of the front  and back MWPC
respectively.  These  reference wires are  used to align  the detector
assembly   and   obtain   the   position   resolution   (see   Section
\ref{sec:PosAnalysis}).   The   detector  system  is   operated  using
isobutane $i$(C$_4$H$_{10}$), with gas  pressure ranging between 2 and
6~mbar.  Voltages on the cathode  are chosen to optimize the amplitude
of the signal (depending on the dynamic ranges for the energy loss and
velocity  of  the  detected  ions).  Typical  values  ranging  between
$-415$~V and  $-475$~V, were used.  Permanent  magnets, placed outside
the reaction chamber, were sufficient to suppress the remaining effect
of $\delta$-electrons  originating from the atomic  interaction of the
heavy-ions with the target.

The MWPC  timing signals  were amplified by  ORTEC FTA820  fast timing
amplifiers.   The  amplified signals  were  sent  to  an Enertec  7174
Constant  Fraction  Discriminator (CFD)  using  20\%  of the  original
signal,  that  provided  the  start  signals  for  the  time-of-flight
measurements.    The  stop  signal,   generated  using   an  analogous
electronic chain,  was provided  by a large  area MWPC located  in the
focal  plane of  {\sc VAMOS}++~\cite{Rej11}.   The  time-of-flight was
measured using  an ORTEC 566  Time to Amplitude Converter  (TAC).  The
charge  collected  on  each   wire  was  individually  integrated  and
multiplexed     using      GAS     SIlicium     multiPLEXing     chips
(GASSIPLEX)~\cite{San94}.   The  readout  was  ensured  by  CAEN  V551
sequencer  and CAEN  Readout for  Analog Multiplexed  Signals (C-RAMS)
V550 modules.

The DPS-MWPC at {\sc VAMOS}++ has the following advantages:
\begin{enumerate}[(i)]
\item Direct and precise  two-dimensional measurements of the position
  of  reaction   products  provide   the  scattering  angle   and  the
  interaction point on the target. This also allows the control of the
  shape and position  of the beam spot.  The  twofold measurement also
  allows to increase the detection efficiency~\cite{Kumagai2013}.
\item A pair of fast timing signals are available for the start of the
  time-of-flight measurements.
\item  Individual readout  of  charges collected  on  the wire  planes
  allows a  uniform signal to  noise ratio over the  complete detector
  size      as     compared      to     the      use      of     delay
  lines~\cite{Beghini2005,Kumagai2013,Jhin14}.   Further,  the  direct
  measurement  of   the  charge  distribution   allows  a  multi-track
  detection  capability~\cite{Breskin1982} and  thus the  treatment of
  pile-up events.
\item  The  use  of a  common  volume  of  gas,  at low  gas  pressure
  (2-6~mbar),  with  thin  Mylar  windows (0.9$\mu$m),  results  in  a
  relatively  small energy  loss  and angular  straggling  that has  a
  minimum impact on the  performance of the magnetic spectrometer and
  focal plane detection system.
\item A limited sensitivity to  the magnetic fringe fields (from large
  aperture quadrupoles) allows an operation without the constraints of
  magnetic shielding.
\item  A limited  sensitivity  to $\delta$-electrons  produced in  the
  interaction of heavy  ion beams with the target  allows an operation
  at higher beam intensities in the absence of electrostatic mirrors.
\end{enumerate}

\begin{figure*}
\begin{center}
\includegraphics[width=\textwidth]{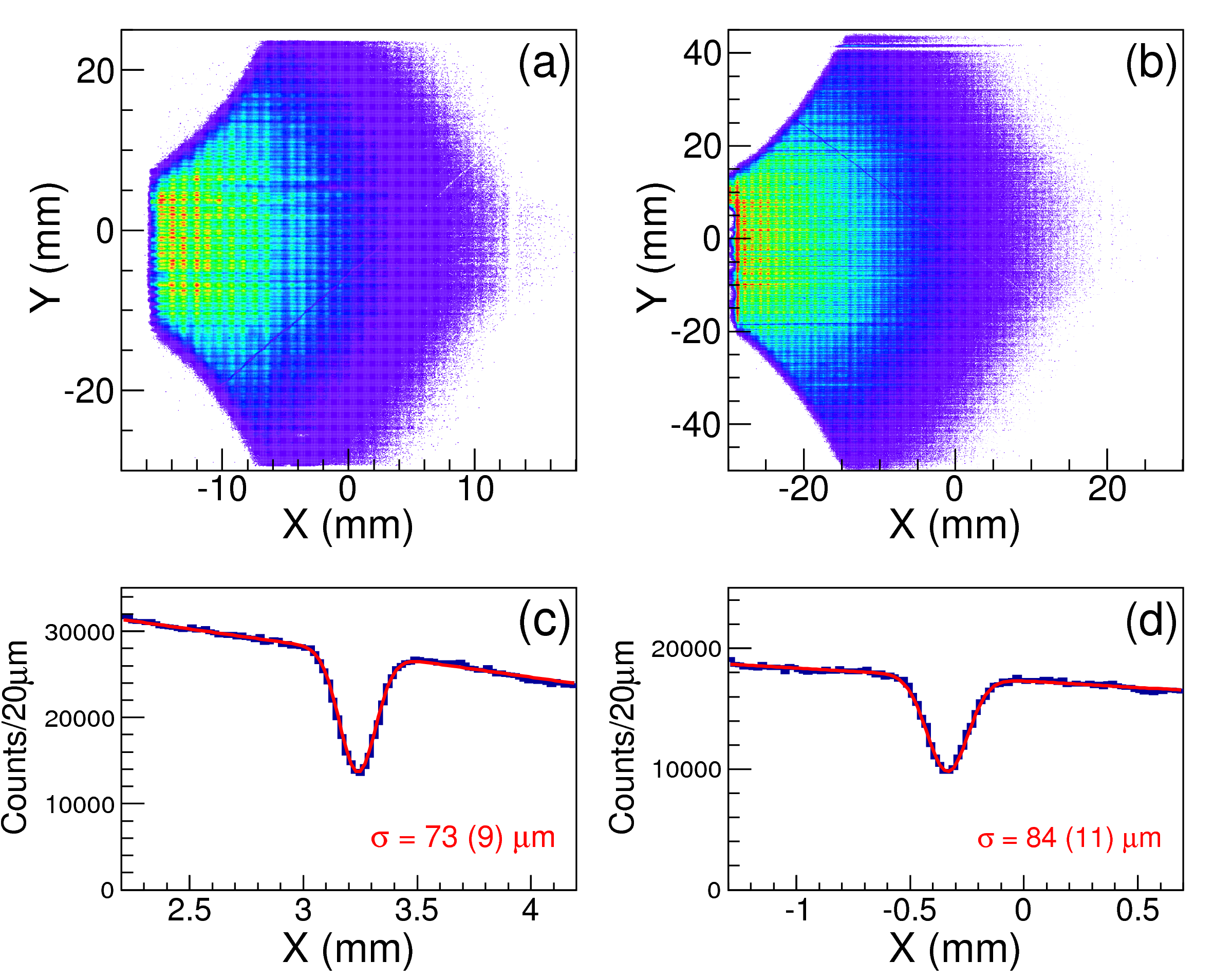}
\caption{ (Color online) Projected positions on the front (a) and back
  (b) reference  wire planes.  (c) The reconstructed  events of figure
  (a) are  projected on a  plane perpendicular to the  front reference
  wire. (d) same as figure (c) for the back reference wire.  The solid
  line show  a fit with the  function $D(x)$ (see  text). The position
  resolution is also indicated.
\label{fig:FigRecPos}}
\end{center}
\end{figure*}
\section{Detector performance}
\label{Sec:Perfs} 
In this  section, the performance  of the DPS-MWPC assembly,  based on
source  and in-beam  measurements made  at GANIL,  is  presented.  The
time-of-flight, position  and angular resolutions of  the new detector
reported in this section are summarized in Tab.~\ref{tab:perf}.

\subsection{Measurements}
\label{Sec:PerfsExp} 
Two independent  measurements were performed to  quantify the position
and timing  performances of the detector.  (a)  Measurements were made
using a collimated $^{252}$Cf fission  source placed 10~cm in front of
the detector assembly.  For  these measurements an additional MWPC was
placed 2~cm in front of the DPS-MWPC and a silicon detector was placed
4~cm behind.   The detector was  operated at $2.3$~mbar and  a cathode
voltage of $-430$~V.   (b) An in-beam measurement was  also made using
$^{238}$U  beam   at  6.2~MeV/u   ($\sim  0.2$~pnA)  impinging   on  a
1.85~mg/cm$^{2}$ thick $^{9}$Be target. The fragments from fusion- and
transfer-  induced  fission   were  isotopically  identified  in  {\sc
  VAMOS}++ placed at $26^\circ$. The detector was operated at $6$~mbar
of isobutane, and $-470$~V was applied on the cathodes.

\begin{figure*}[t]
\includegraphics[width=\textwidth]{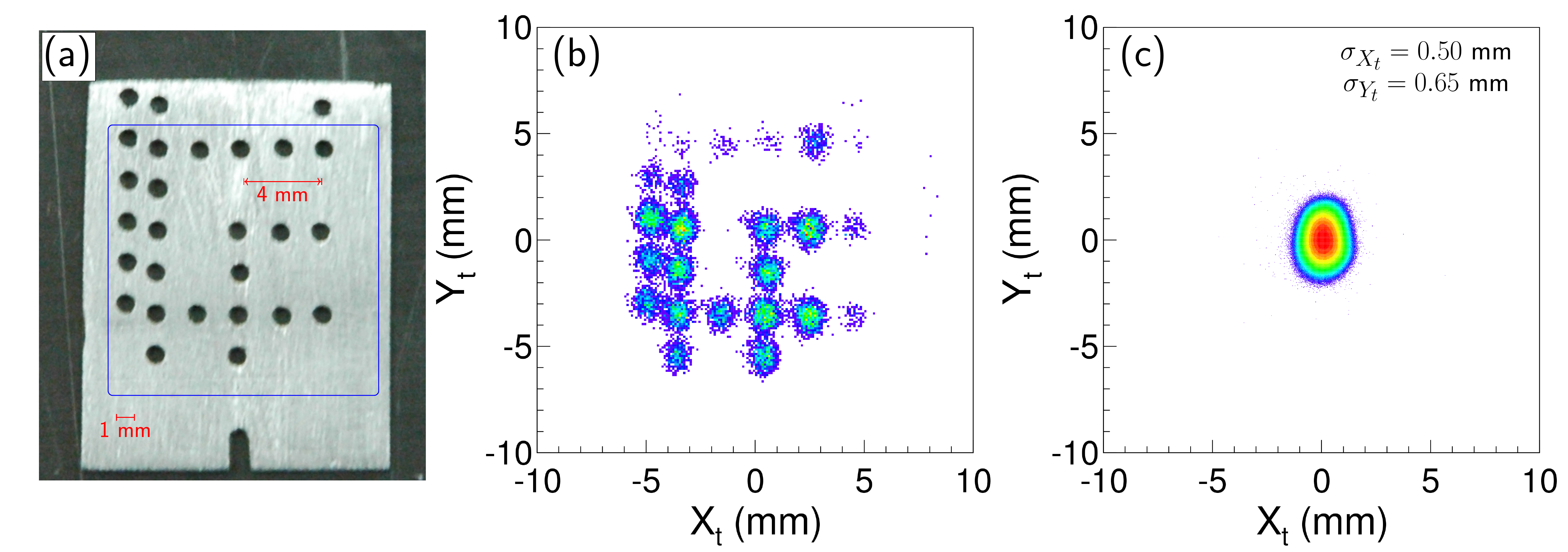}
\caption{\label{fig:FigMask} (Color online) Position reconstruction at
  the target position.   (a) Mask placed in front  of the target.  The
  holes  have  a  diameter  of  1~mm  with  different  spacings.   (b)
  Reconstructed image of the  mask.  (c) Measured profile of $^{238}$U
  beam. Standard deviations $\sigma_{X_{t},Y_{t}}$ are indicated.}
\end{figure*}

\subsection{Position resolution}
\label{sec:PosAnalysis}
For each  MWPC, an  event is  characterized by a  timing signal  and a
charge  distribution   in  the  $X$  and  $Y$   wire  planes.   Charge
distributions with a typical  wire multiplicity of five were observed.
The gain matched charge distributions,  with a wire multiplicity of at
least three, were  used to extract the position.   A hyperbolic secant
function~\cite{Lau1995}   was   fitted    to   the   measured   charge
distributions to obtain the positions ($X_1$,$Y_1$), ($X_2$,$Y_2$) for
the front and back MWPC respectively.

The  measurements presented  in this  section were  obtained  from the
in-beam  experiment.  The  detector was  operated in  coincidence with
ions    detected   in   {\sc    VAMOS}++.    The    four   coordinates
($X_1$,$Y_1$,$X_2$,$Y_2$)  were  used  to reconstruct  the  scattering
angles    ($\theta_V$,$\phi_V$)    of    the   detected    ion    (see
Fig.~\ref{fig:fig1}).  Using these coordinates, the projected position
($X$,$Y$)  of  the  ion  trajectory  on any  required  plane,  and  in
particular    at   the   target    position,   could    be   obtained.
Figure~\ref{fig:FigRecPos}(a)  and (b) show  these projections  on the
plane  of the front  and back  reference wires.   The envelope  of the
distribution of events can be  seen to trace the angular acceptance of
the spectrometer~\cite{Pul08}.  The  intensity distribution is related
to the  kinematics of  fission (induced by  heavy-ion beam  in inverse
kinematics).  The shadows of the reference wires (100~$\mu$m diameter)
can also  be seen.  The reduction  in the number of  events at regular
intervals (0.5~mm and  1.0~mm along Y and X  axis respectively) can be
seen in  Fig.~\ref{fig:FigRecPos}.  This arises from  the shadowing by
the  cathode and anode  wires.  The  two-dimensional spectra  shown in
Fig.~\ref{fig:FigRecPos}(a)  and  (b)  can  be projected  on  a  plane
perpendicular   to  the   corresponding   reference  wires.    Figures
\ref{fig:FigRecPos}(c) and  (d) show  these projections for  the front
and back reference  wires respectively.  The profile of  the shadow of
the wire is  used to measure the position  resolution of the detector.
The above  projected spectra  were fitted using  a function that  is a
convolution  of a  slit and  a Gaussian  function. This  describes the
shadowing  of  a  wire  with  given diameter  superimposed  on  linear
distribution as a function of position and has the following form:

\begin{equation}
  \nonumber
  D(x)\rm{=}\left(1\rm{-}\frac{1}{2}\left(\Phi\left(\frac{x\rm{-}\frac{d}{2}}{\sqrt{2}\sigma}\right)+\Phi\left(-\frac{x\rm{+}\frac{d}{2}}{\sqrt{2}\sigma}\right)\right)\right)\rm{\times}\left(a
  \rm{+} b x\right)
\end{equation}
where:   ${\Phi(x)  =   \frac{1}{2}  \left(1-erf(x)\right)}$   is  the
convolution  of  a step  and  a  Gaussian  function, $\sigma$  is  the
standard deviation describing the position resolution, $d$ is the wire
diameter, $a$  and $b$  parameters are used  to locally  reproduce the
linear behaviour  as function of $x$.   Using a fixed  diameter of the
wire ($d=100~\mu$m), an optimization  of the measured projection using
this  function  results  in  resolutions  of  $\sigma=73(9)~\mu$m  and
$\sigma=84(11)~\mu$m   for  the   front  and   back   reference  wire,
respectively.   The error on  the resolutions  arises mainly  from the
wire deformation  ($\sim~8~\mu$m and $\sim~10~\mu$m for  the front and
back MWPC  respectively).  This was estimated by  using the projection
for various sub-parts of the reference wire.  Additional contributions
arise   from  the  uncertainties   on  the   diameter  of   the  wires
($\sim~3~\mu$m), the position  of the projection plane ($\sim~1~\mu$m)
and of statistical nature ($\sim~1~\mu$m).  As the projection involves
all the  ($X_1$,$Y_1$,$X_2$,$Y_2$) coordinates, the  quoted resolution
can be considered as an upper  limit of the intrinsic resolution for a
single detector.

\begin{figure}[t]
\includegraphics[width=\columnwidth]{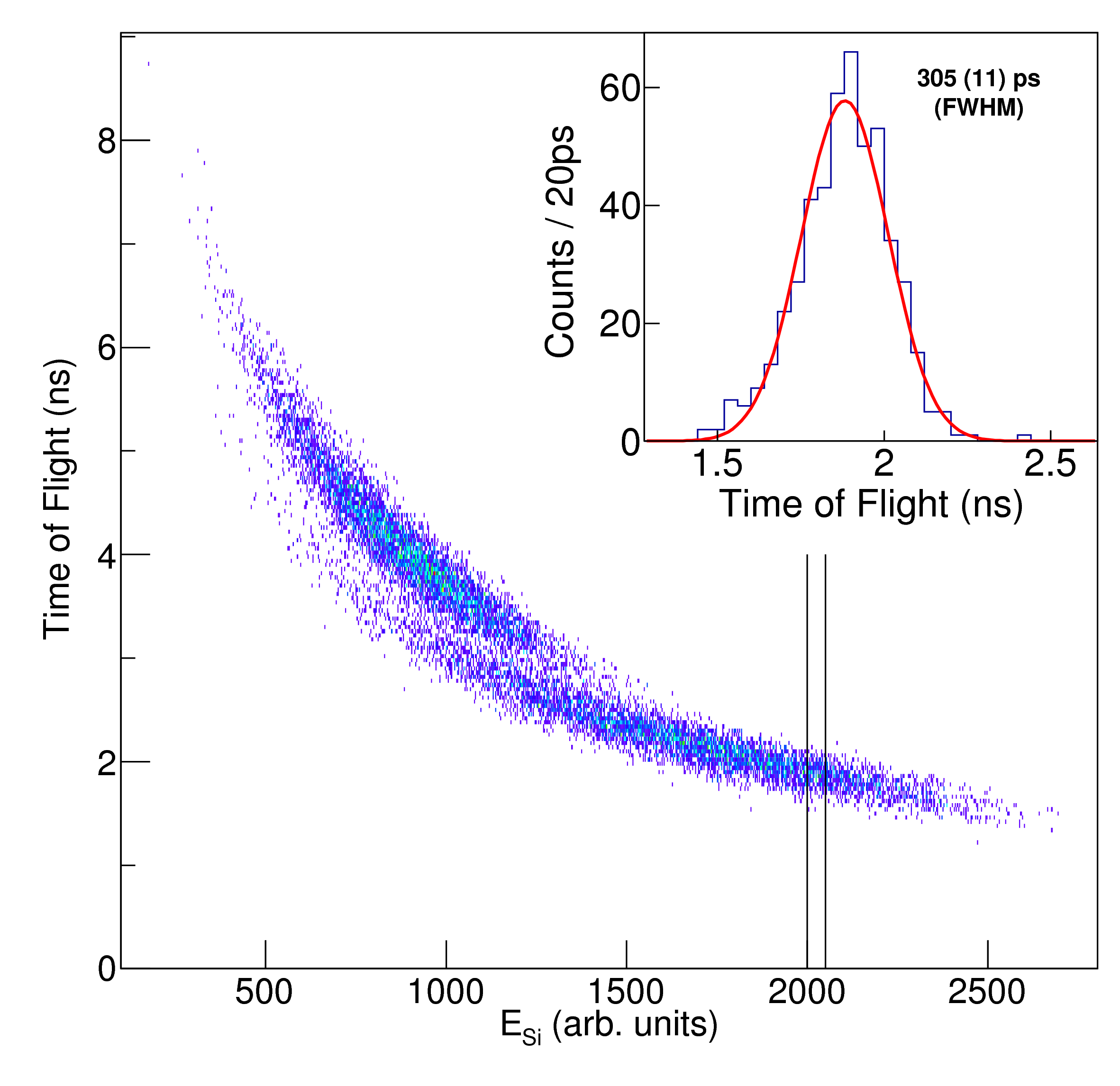}
\caption{\label{fig:TRes} (Color online) Bi-dimensional spectra of the
  time-of-flight  between  reference  and  front  MWPC  versus  the
  residual energy  measured in Silicon  detector. The heavy  and light
  fragments from fission of $^{252}$Cf are separated.  The inset shows
  the  time-of-flight  distribution of  events  in  the energy  region
  depicted by the  two solid lines. The red  line shows the adjustment
  by a Gaussian function with 305~(11)~ps (FWHM).}
\end{figure}
The   resulting   angular   resolution   of  the   scattering   angles
($\theta_V$,$\phi_V$) and position resolution of the interaction point
on  the target  ($X_t$,$Y_t$) were  also investigated.   A  1~mm thick
aluminum mask,  with 1~mm diameter holes,  was placed in  front of the
target          (shown          in         Fig.~\ref{fig:FigMask}(a)).
Figure~\ref{fig:FigMask}(b)  shows   the  corresponding  reconstructed
position using  the detected fragment.  To evaluate  the resolution of
the projected position and scattering angles, a Monte Carlo simulation
for scattered ions,  emitted from a 1\,mm diameter  hole at the target
position, was  performed.  The measured  position resolutions included
in the  simulation reproduced the measured reconstructed  image at the
target  position  (Fig.~\ref{fig:FigMask}(b)).   A  resulting  angular
resolution  (FWHM) for  $\theta_V$ and  $\phi_V$ was  evaluated  to be
$2.5(3)$~mrad ($\sim 0.14^\circ$).  Similarly, the position resolution
(FWHM)  of the interaction  point on  the target  was evaluated  to be
$560(70)$~$\mu$m.   Figure~\ref{fig:FigMask}(c) shows  the  profile of
the  $^{238}$U  beam  measured  with the  DPS-MWPC,  illustrating  the
tracking capabilities of the detector.

\subsection{Time resolution}
The time-of-flight  resolution of the  MWPCs was determined  using the
$^{252}$Cf  source.   The  time-of-flight  was  measured  between  the
additional MWPC  and the front  MWPC.  The residual  energy (E$_{Si}$)
was  measured in  the silicon  detector.  The  two-dimensional spectra
shown in  Fig.~\ref{fig:TRes} illustrates the  correlation between the
time-of-flight  and the energy.   As can  be seen  in the  figure, the
light and heavy fission  fragments are well separated.  The resolution
(FWHM)  for the  time-of-flight,  obtained using  an energy  selection
(illustrated  in  Fig.~\ref{fig:TRes})  on  a fraction  of  the  light
fission fragments, was measured to be $305(11)$~ps.  Assuming that the
start and  stop detector have similar time  resolutions, the intrinsic
time resolution (FWHM) is estimated to be $\sim 216~$ps.

\subsection{Efficiency}
The transparency of each MWPC was estimated to be 96\% considering the
loss of ions due to the stopping by the 20~$\mu$m wires of the cathode
and  anode planes.   The detection  efficiency was  obtained  from the
in-beam  measurement for  heavy  ions with  $28  \leq Z  \leq 65$  and
energies ranging between 2 to 8~MeV/u.  The trigger of the acquisition
was obtained  from {\sc VAMOS}++  detecting a fission  product.  Using
the correlation  of the  relevant signals from  {\sc VAMOS}++  and the
timing signal  and positions ($X$,$Y$) from the  DPS-MWPC detector, an
efficiency $\sim 98$\% of the reconstructed parameters ($t$, $X$, $Y$)
for each  MWPC plane was obtained.  The  efficiency for reconstructing
an event ($t$, $X_t$, $Y_t$,  $\theta_V$, $\phi_V$) was measured to be
$\sim 96$\%.

\begin{table}
\caption{Summary    of    the     performance    of    the    DPS-MWPC
  detector.\label{tab:perf}} \centering
\begin{tabular}{c|c|c|c}
 Quantity & Description & Unit & $\sigma$  \\
 \hline
   $t_1$ &Time-of-flight & ps & $130(5)$  \\
   $X_1$, $Y_1$ &Front MWPC & $\mu$m & $73(9)$ \\
   $X_{2}$, $Y_{2}$ &Back MWPC & $\mu$m &  $84(11)$  \\
   $\theta_V$, $\phi_V$& Scattering angles & mrad & $1.1(1)$ \\
   $X_t$, $Y_t$ & Interaction  position  & $\mu$m & $239(30)$  \\
 & on the target  & 
 \end{tabular}
\end{table}

\subsection{Doppler correction of $\gamma$-ray energies}
\begin{figure}[t]
\begin{center}
  \includegraphics[width=\columnwidth]{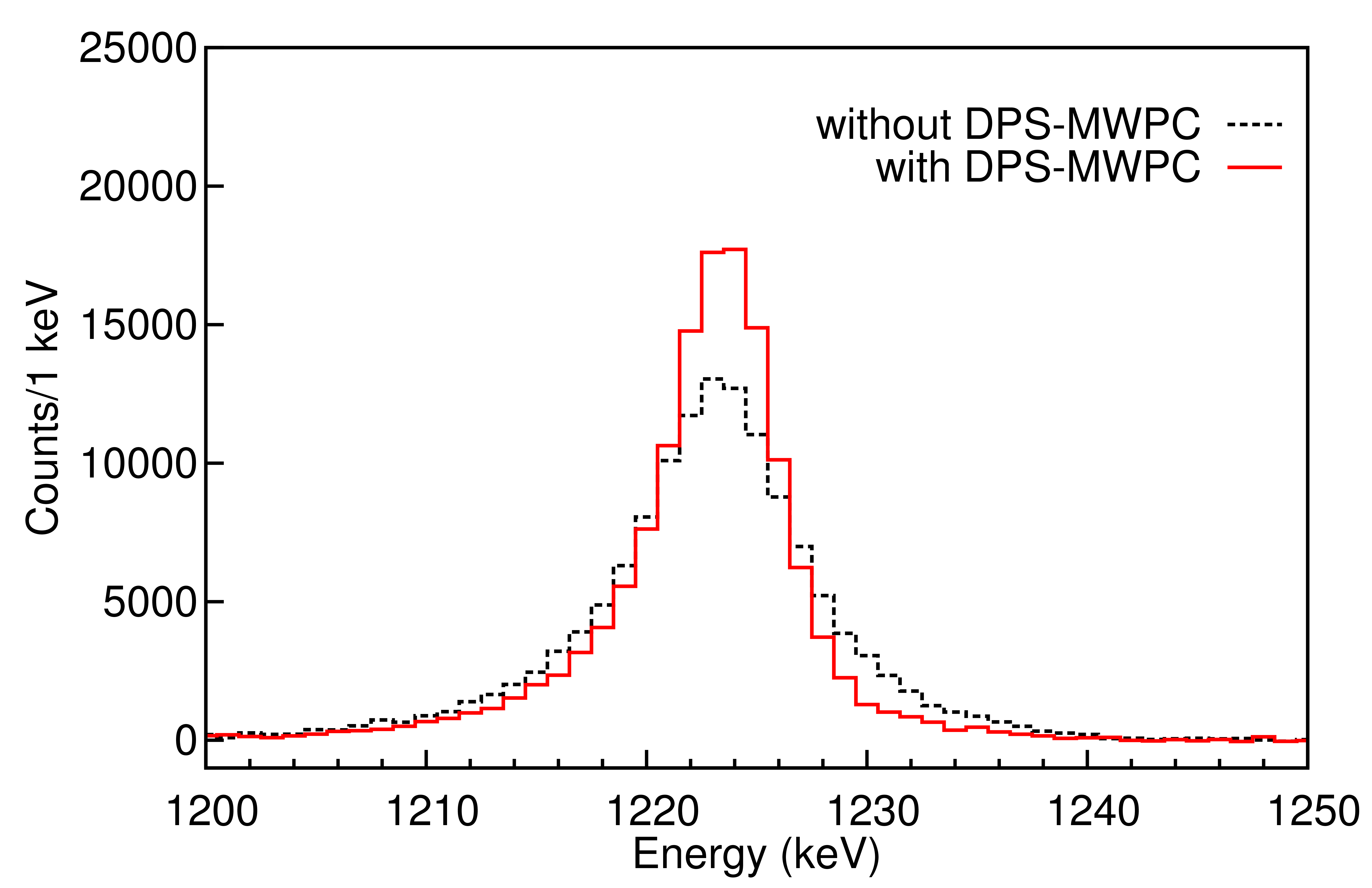}
  \caption{\label{fig:FigGamma}  (Color online)  Part  of the  tracked
    event-by-event  Doppler  corrected  $\gamma$-ray  spectra  in  the
    region of the $1222.9(1)$~keV transition, measured in {\sc AGATA},
    in coincidence with isotopically  identified $^{98}$Zr in the {\sc
      VAMOS}++  spectrometer.   Red  solid  line: using  the  measured
    $^{98}$Zr velocity vector and interaction point on target with the
    DPS-MWPC detector.  Black dotted  line: using angles at the target
    obtained from  the ion trajectory reconstruction  method from {\sc
      VAMOS}++ (see text).  }
\end{center}
\end{figure}

The   performance  of   the  event-by-event   Doppler   correction  of
$\gamma$-ray  energies   using  the  new  DPS-MWPC   detector  is  now
discussed.   Fission  fragments   from  fusion-  and  transfer-fission
produced in the collision of the $^{238}$U beam with the $^9$Be target
were   detected   and  isotopically   identified   in  {\sc   VAMOS}++
spectrometer,  (see Sect.\ref{Sec:PerfsExp}).  Their  velocity vectors
were   measured  using   the   DPS-MWPC  detector   as  described   in
Sect.~\ref{Sec:Perfs}.  The {\sc AGATA}~\cite{AGATA} detectors covered
angles from 100$^\circ$ to 170$^\circ$.   The {\sc AGATA} array was in
a compact  configuration (translated by 8.8~cm  downstream compared to
the nominal  configuration, where the  detectors are at a  distance of
23.5~cm  from  the  target).   The  $\gamma$-ray  emission  angle  was
determined  using its  first  interaction point,  obtained from  pulse
shape  analysis  and   tracking  procedures~\cite{AGATA}.   A  typical
position  resolution  of  5~mm~(FWHM)~\cite{Recchia2008,  Recchia2009,
  Soderstrom2011} has  been reported for  $\gamma$-ray energies around
1.3~MeV.   This  corresponds  to   an  angular  uncertainty  of  $\sim
1.9^\circ$  in  a  compact  configuration.   Figure~\ref{fig:FigGamma}
shows part of the Doppler corrected $\gamma$-ray spectra including the
known  1222.9~(1)~keV  transition   measured  using  {\sc  AGATA},  in
coincidence  with $^{98}$Zr fragment  isotopically identified  in {\sc
  VAMOS}++.  The  solid line corresponds to  an event-by-event Doppler
correction using  the measured velocity  vector of the  scattered ion,
the derived interaction point on  the target and the first interaction
point of  the $\gamma$-ray  in {\sc AGATA}.   The dashed  line spectra
corresponds to the  Doppler correction where the DPS-MWPC  is not used
and  the  scattering  angles  were  derived using  an  ion  trajectory
reconstruction  method~\cite{Rej11,  Pul08,  Pul08-2}.   The  improved
resolution resulting  from the direct measurement  of scattering angle
and  the derived  position  of the  reaction  point at  the target  is
evident from  the figure.  The resolution  is improved by  a factor of
$\sim 1.6$ (5.5~(1)~keV compared  to 8.6~(1)~keV).  It should be noted
that in the  nominal configuration, where the detectors  are placed at
$23.5$~cm  from  target,  a   smaller  angular  uncertainty  of  $\sim
1.2^\circ$ will lead to  a superior Doppler correction.  The excellent
performances  of the DPS-MWPC  reported here,  with a  typical angular
resolution  of  $\sim 0.14^\circ$~(FWHM)  and  the  resolution of  the
derived    position    of     interaction    at    the    target    of
$\sim560~\mu$m~(FWHM), are approximately one order of magnitude better
than that required for {\sc AGATA} and hence is compatible with future
improvements in the $\gamma$-ray tracking performances.

\section{Summary and perspectives}
The characteristics  and performance of a new  dual position sensitive
MWPC  detector system  are reported.   This  detector is  used at  the
entrance  of the  {\sc VAMOS}++  spectrometer and  provides  both fast
timing signals and two-dimensional position coordinates.  These define
the trajectory  of the reaction products, namely  the scattering angle
and interaction  point on the target.  A  time-of-flight resolution of
$305(11)$~ps  (FWHM)  and a  position  resolution of  $172(21)$~$\mu$m
(FWHM) have  been measured.   Angular resolution of  $2.5(3)$~mrad and
resolution of the interaction  point at the target of $560(70)$~$\mu$m
were obtained.  An overall efficiency  of $\sim 96\%$ was measured for
fission fragments over a wide range of velocities.  The above tracking
performances were  applied to  the Doppler correction  of $\gamma$-ray
energies with  the $\gamma$-ray tracking array  AGATA.  The resolution
in the Doppler corrected  $\gamma$-ray energy at $1.2$~MeV is improved
by a  factor of $\sim 1.6$  compared to the result  obtained using the
previous ion trajectory reconstruction method.  The angular resolution
of the  DPS-MWPC is approximately  one order of magnitude  better than
what is required today for a $\gamma$-ray tracking array.  It can thus
cope  with   the  improvements  in  the  angular   resolution  of  the
$\gamma$-ray detection systems.

The DPS-MWPC  was routinely used for  beam tuning and  to optimize the
spectrometer performance  during the first  campaign of AGATA  at {\sc
  VAMOS}++ in  2015.  In the  compact configuration of the  AGATA, the
event-by-event measurement  of the interaction point on  the target is
important to ensure  the stability of the beam  spot size and position
throughout the experiment.  For experiments employing direct reactions
relying        on         the        kinematical        energy-angular
correlations~\cite{Catford2010,Caamano2015}, the precise determination
of the interaction point at the target is essential.

Presently, the performances of  the large acceptance spectrometer {\sc
  VAMOS}++ rely  on an ion trajectory reconstruction  method.  In such
an approach, the measurement of ion trajectories at the focal plane is
used to derive the magnetic rigidity, path length in the spectrometer,
and the  scattering angles at the  target, assuming a  point like beam
spot. It  has been  shown in  the present work  that the  new detector
leads  to  a large  improvement  in  the  determination of  scattering
angles.  This  precise determination of  the scattering angles  and of
the  interaction point on  the target,  obtained independently  of the
spectrometer,  coupled  to the  information  from  {\sc VAMOS}++  will
result  in further  improvements  of the  resolution  in the  magnetic
rigidity and path length and thus provide an improved mass resolution.

Tracking and timing information at the focal plane of {\sc VAMOS}++ is
presently  provided by  two drift  chambers  and one  large area  MWPC
respectively. The  detector reported in this paper  represents a first
step towards  an implementation of large area  dual position sensitive
MWPC for  both tracking and  timing at the  focal plane of  VAMOS that
will  result  in  lower  energy  losses  and  improved  counting  rate
capabilities.

\section*{Acknowledgements}
We would like to thank the AGATA collaboration for the availability of
the  AGATA  $\gamma$-ray  tracking  array  at GANIL.   We  also  thank
G.~Duch\^ene and the E680 collaboration for providing the relevant raw
data  for  Fig.~\ref{fig:FigGamma}.    We  acknowledge  the  important
technical  contributions  of  L.~M\'enager,  J.~Ropert and  the  GANIL
accelerator staff.

\section*{}

\bibliographystyle{elsarticle-num2} 
\bibliography{BiblioMW}

\end{document}